Branched chain amino acids, an "essential" link between diet, clock and sleep?

Laurent Seugnet

Centre de Recherche en Neurosciences de Lyon, Equipe Physiologie intégrée du système d'éveil (WAKING), Université Claude Bernard Lyon 1, INSERM U1028, CNRS UMR 5292, Bron, France

**Abbreviations**

BCAA : Branched chain amino acids

AAR: Amino Acid Response

Atf5: Activating Transcription Factor 5

BCAT: Branched-Chained Amino Acids Transaminase

BCKD: Branched-Chained alpha-Keto acid Dehydrogenase

BLOC-1: Biogenesis of Lysosome-related organelles Complex-1

EAAT-1: Excitatory Amino Acid Transporter -1

GABA: Gamma Amino-Butyric Acid

GABAT: GABA Transaminase

GCN2: General Control Non-derepressible kinase 2

JhI-21: Juvenile hormone Inducible -21

KO: Knock-Out

LAT-1: L-type Amino acid Transporter - 1

LAT-2: L-type Amino acid Transporter - 2

SLC7A5: Solute Carrier 7A5

SLC3A2: Solute Carrier 3A2

TOR: Target of Rapamycin kinase

Abstract

The branched-chain amino acids: leucine, isoleucine and valine occupy a special place among the essential amino acids because of their importance not only in the structure of proteins but also in general and cerebral metabolism. Among the first amino acids absorbed after food intake, they play a major role in the regulation of protein synthesis and insulin secretion. They are involved in the modulation of brain uptake of monoamine precursors with which they may compete for occupancy of a common transporter. In the brain, branched-chain amino acids are involved not only in protein synthesis but also in the metabolic cycles of GABA and Glutamate, and in energy metabolism. In particular, they can affect GABAergic neurons and the excitation/inhibition balance. Branched-chain amino acids are known for the 24-hour rhythmicity of their plasma concentrations, which is remarkably conserved in rodent models. This rhythmicity is partly circadian, independent of sleep and food.  Moreover, their concentration increases when sleep is disturbed and in obesity and diabetes. The mechanisms regulating these rhythms and their physiological impact remain poorly understood. In this context, the Drosophila model has not yet been widely used, but it is highly relevant and the first results indicate that it can generate new concepts. The elucidation of the metabolism and fluxes of branched-chain amino acids is beginning to shed light on the mysterious connections between clock, sleep, and metabolism, opening the possibility of new therapies.

Keywords: branched-chain amino acids, metabolism, GABA, sleep, circadian clock, drosophila

Introduction

The sleep-wake cycle is a highly integrative process for maintaining optimal brain function and, directly or indirectly, the health of the whole body, including metabolic health. In addition, maintaining the intense metabolic activity of the brain requires a substantial supply of nutrients compared to other organs. This is reflected in the strong interactions between regulation of the sleep/wake cycle and dietary need, as well as in the metabolic disturbances associated with sleep disturbances [1-3]. To understand these interactions and their consequences it is absolutely necessary to elucidate the metabolic regulations that are at stake. In this context, amino acids, and in particular proteinogenic amino acids, play a pivotal role: 1) they are necessary for the synthesis of direct precursors of neurotransmitters (monoamines, GABA), or are themselves major brain neurotransmitters (glutamate, glycine, etc.). ); 2) they are the building blocks for protein synthesis; 3) they are anaplerotically involved in energy metabolism; 4) unlike other nutrients, they cannot accumulate in intracellular storage systems such as glycogen and lipid droplets; 5) essential amino acids can only be supplied by diet or protein degradation and autophagy.

Among the 9 essential proteinogenic amino acids in mammals, the 3 branched-chain amino acids (BCAA): leucine, isoleucine and valine, constitute a particular group. They are abundant in proteins, representing about 20% of the amino acids, and 40% of the minimum necessary intake of essential amino acids [4]. Leucine is thus the most frequent amino acid in protein sequences, with a frequency close to 10%, both in eukaryotes and prokaryotes [5]. Biochemically, all three BCAAs are hydrophobic amino acids, required for the conformational folding of globular proteins, but despite this commonality, they are not easily interchangeable in a protein sequence. Leucine plays a critical role in protein-protein interactions more frequently than the other BCAAs. This is particularly the case in the leucine zipper, a motif that allows the dimerization of many transcription factors including c-fos and c-jun [4], where the leucines are regularly spaced, and aligned in the dimer. It has been proposed that this functional importance contributes to the fact that it has six codons associated with it, the largest number for an amino acid (along with arginine), thus further protecting leucine from genetic point mutations that can affect coding sequences [6]. The prominent role of BCAAs and leucine in particular is also reflected in their ability to regulate insulin secretion and protein synthesis. BCAAs are also well known for their involvement in the synthesis of neurotransmitters, their association with circadian rhythms and sleep disorders. Before detailing this association and to try to better interpret it, we will recall the peculiarities of BCAAs transport and metabolism. The objective of this update is to put into perspective the current research in relation to the accumulated knowledge on this topic, without being exhaustive, to stimulate reflection.

Transport of free BCAAs

Although plasma free amino acids represent a small fraction of the total amino acid pool in the human body (estimated at 0.5%), their metabolic role is crucial [7]. Following protein intake, essential amino acid concentrations increase substantially, and BCAAs are among those whose concentration increases most rapidly and proportionally to protein intake [8-11] (Figure 1). The fact that the liver does not directly metabolize BCAAs contributes to their rapid increase in the circulation after absorption, which is not the case for other amino acids and in particular for small non-essential amino acids present in high concentration in plasma such as glycine, alanine and glutamine. This increase in BCAAs, especially leucine, contributes to insulin secretion by pancreatic β-cells through two identified mechanisms: 1) the direct metabolite of leucine transamination, α-ceto-isocaproate acid, increases ATP production at the mitochondria, and thus activation of ATP-dependent potassium channels that can induce insulin release, 2) leucine can allosterically activate glutamate dehydrogenase and thereby increase intracellular citrate and energy metabolism, which is a signal for insulin release [12]. Circulating insulin in turn promotes the absorption of amino acids [12,13].

Leucine and to a lesser extent other circulating BCAAs also stimulate protein synthesis [14]. This effect depends on signaling through the Target of Rapamycin (TOR) kinase and other TOR-independent pathways [12]. A low protein diet reduces plasma BCAA concentrations [15]. In contrast, hypoglycemia related to hypocaloric feeding or fasting (as early as 24 h) is associated with an increase in circulating amino acids, particularly BCAAs [15,16]. During fasting, the increase in circulating BCAAs is largely due to protein autodigestion by muscle cells, which contain the largest pool of protein in the body [4,15]. In the absence of glucose, two metabolic reactions common to the three BCAAs lead to their degradation: transamination by BCAA transaminase (BCAT) followed by oxidation of the branched keto acids by branched keto acid dehydrogenase (BCKD). This last reaction is irreversible and allows the energy metabolism to be fed by ketogenesis and neoglucogenesis.

The entry of plasma BCAAs into the brain depends on their transport through the blood-brain interfaces: the blood capillaries and the choroid plexuses. This transport involves more than 20 different transporters that can be classified into two main categories: "loaders" and "harmonizers" [17]. Schematically, loaders, such as the glutamate transporter EAAT-1, mainly use the sodium gradient for energy-dependent transport of non-essential amino acids, inducing concentration differentials between intra- and extracellular environments. On the other hand, harmonizers have a wide selectivity and facilitate the transport of solutes along their concentration gradients. Most harmonizers are antiporter systems: the import of an amino acid is conditioned by the export of another amino acid. This makes it possible to exploit the gradients generated by the loaders to import plasma amino acids, especially essential amino acids, while maintaining a balance of concentrations. Indeed, free amino acids are generally more concentrated in the intracellular medium than in plasma, and their concentration varies substantially depending on the type of amino acid considered [18-21]. In both plasma and cells, small amino acids, which can be synthesized by the cell and serve as metabolic intermediates, are more concentrated. With organ-specific variations, this type of functioning applies to all cells in the body and is not specific to the brain. At the blood-brain barrier, the import of BCAAs and more generally of large neutral amino acids relies largely on the heterodimeric transporter LAT-1 (L-Type Amino acid Transporter - 1 or SLC7A5, combined with the heavy chain SLC3A2) which is expressed at a very high level (100-fold higher than in the periphery) in the luminal and abluminal membranes of cerebral capillary endothelial cells [22] (Figure 1). LAT-1 is an antiporter, and in tumor cells where it is also abundantly expressed, it has been proposed that LAT-1 exports glutamine exchange for leucine, allowing activation of TOR signaling and cell growth [23]. In the case of blood-brain interfaces, glutamine is abundantly exported and allows the import of neutral amino acids [13] but the intracellular substrate(s) of LAT-1 in this context have not been clearly identified [24]. Leucine, which together with tyrosine and phenylalanine is the most rapidly imported amino acid into the brain [25], is clearly dependent on LAT-1 for its uptake. Conditional inactivation of LAT-1 in endothelial barrier cells induces a brain deficit of leucine and isoleucine, as well as an excess of histidine, confirming the importance of this transporter for the import of BCAAs, and suggesting that histidine could be the substrate exported in this context [26] (see also [23,27]). LAT-1 is not present in the choroid plexus, but a paralog, LAT-2, is highly expressed there and fulfills a role complementary to that of LAT-1 in the regulation of BCAA and glutamine [24,28]. An important feature of neutral amino acid transport at the barrier is its saturation under normal conditions with a Km equal to or lower than plasma concentrations, which means that there is competition between different amino acids to be taken up by brain endothelial cells [29]. At physiological plasma concentrations, leucine and phenylalanine alone would account for 50% of LAT-1 occupancy [25,30]. A change in the concentration of these amino acids can therefore have a major impact on transport at the barrier. Thus, reducing plasma BCAA concentrations in mice deficient for BCKDK, the kinase that phosphorylates and inhibits BCKD, involved in BCAA degradation (see above), also reduces BCAA concentrations in the brain but those of phenylalanine, histidine, threonine and methionine are significantly increased by up to 200% [31]. Such competition does not exist in peripheral tissues where the transport systems have a lower affinity, with a Km that is at least an order of magnitude above plasma concentrations [29]. This saturation of transport at the barrier and competition between substrates may lead to changes in the import of amino acids that are precursors of monoamines. It has been noted for decades that the import of tryptophan in particular, one of the least present amino acids in plasma, is conditioned by the plasma concentrations of other neutral amino acids. Tryptophan being the limiting precursor for serotonin synthesis, via tryptophan β-hydroxylase, any manipulation impacting the competition with other amino acids is able to positively or negatively modulate brain serotonin levels. Tyrosine transport, which is also dependent on LAT-1, similarly affects catecholamine synthesis [8,32]. BCAA supplementation can thus simultaneously reduce the levels of serotonin and catecholamines, including dopamine, by inhibiting the import of their respective precursors [33-35]. Given the importance of these neurotransmitters in the regulation of sleep, many strategies based on various combinations of nutrients and amino acids have been tested in order to improve alertness states [36]. The results have been mixed and difficult to reproduce. This lack of reliability is probably due in part to the fact that a moderate increase in the synthesis of a neurotransmitter is not a sufficient condition to intensify neurotransmission, except in certain specific situations. It is possible that tryptophan supplementation could nevertheless improve sleep in certain metabolic contexts related to chronic elevation of circulating BCAA levels such as diabetes and obesity, and tests are still being conducted in this direction [37]. Competition with monoamine precursors for import into the brain is not the only BCAA-dependent process that can influence neurotransmission. Indeed, after their uptake by cells at the blood-brain interfaces, BCAAs are transported at least in part by LAT-1 and LAT-2 into astrocytes and neurons, where they play a decisive role not only in protein synthesis but also in the regulation of the GABA-glutamate-glutamine cycle and in energy metabolism [24].

Metabolism of BCAAs, links with neurotransmitter synthesis

The role of BCAAs in the synthesis of glutamate and GABA is detailed in several reviews (e.g. [38,39]), we will only mention here the major elements. The first step in BCAA metabolism is transamination: branched-chain amino acid transaminase (BCAT), yields glutamate and a keto acid from BCAA and α-ketoglutarate, a Krebs cycle intermediate (Figure 1). This de novo synthesis of glutamate is favored in astrocytes where mitochondrial BCAT is expressed. In these cells glutamate is then converted to glutamine by glutamine synthetase and can thus be integrated into the glutamate-glutamine cycle (see below). Alternatively, glutamine can be exported to import other neutral amino acids via the LAT-1 or LAT-2 antiport systems. Dialysis and metabolic tracing studies suggest a model whereby the ketoacid thus formed is transferred to neurons where cytoplasmic BCAT reconverts it to BCAA and α-ketoglutarate. This reaction uses glutamate, in high concentration in the neurons. The "regenerated" BCAAs can then be transported to the astrocytes in exchange for glutamine, completing a metabolic shuttle. In humans, it is the endothelial cells of the barrier that express mitochondrial BCAT and that could replace astrocytes in this scheme [39,40]. This system completes the glutamate-glutamine cycle by which glutamate released at synapses is recaptured by astrocytes where it is metabolized into glutamine. The latter is then transferred to neurons where it is reconverted into glutamate by glutaminase. It has been estimated that this system only recycles 60-80% of the glutamate, some of which is diverted to other uses. As there is no significant import of plasma glutamate, which protects the brain from potential excitotoxicity, glutamate must be synthesized within the brain parenchyma and in this context BCAAs are a major source of nitrogen. A study performed with isotopically labeled leucine showed that up to 50% of the nitrogen in glutamate can come from leucine [41], but valine and isoleucine are also involved [38,42]. In addition to being a major neurotransmitter, glutamate constitutes, together with glutamine, a limiting precursor for the synthesis of GABA. By feeding these metabolic cycles, BCAAs play a key role in brain metabolism and influence the excitation/inhibition balance. Indeed, cytoplasmic BCAT is prominently expressed in GABAergic neurons [39,43] and single-cell transcriptomic analysis in mice [44] shows preferential expression of SLC7A5 (light subunit of LAT-1), and Atf5, a key molecule in the amino acid response (AAR) pathway, in inhibitory neurons, consistent with a high amino acid requirement in these cells. Furthermore, pharmacological studies indicate that the flow of the GABA-glutamate-glutamine cycle is indeed influenced by the ability of LAT-1 transporters to regulate glutamine [45,46]. However, the ability of GABAergic neurons to release GABA is critically dependent on its synthesis and the presence of glutamine [47]. Consistent with these observations, conditional inactivation of LAT-1 in barrier endothelial cells is associated with reduced numbers of GABAergic vesicles and hyperexcitability in the somatosensory cortex, as well as motor deficits that can be corrected with leucine and isoleucine injections [26]. Furthermore, in a mouse model of brain injury, BCAA concentrations are reduced and BCAA supplementation restores excitability of hippocampal networks and cognitive impairment [48]. BCAA supplementation also improves the antiepileptic effect of the ketogenic diet [49,50].

Apart from GABAergic and glutamatergic neurons, cytoplasmic BCAT expression has also been observed in monoaminergic and peptidergic neurons, especially in the hypothalamus, where it probably participates in other functions [51]. Expression of BCKD, which catalyzes the first irreversible step in BCAA degradation, is also restricted to neurons [52], and is also expressed in capillaries in humans [53]. In neurons, dehydrogenation by BCKD of ketoacids from BCAAs can fuel energy metabolism via the ketogenic pathway (leucine, isoleucine) or the Krebs cycle (isoleucine, valine) but can also participate in the synthesis of other amino acids, cholesterol and lipids [54]. Some neural networks express essential amino acid sensors, including sestrin [55], a leucine sensor, and GCN2 kinase [56], which is involved in the ability to prefer non-BCAA-deficient foods, independently of sensory stimuli.

BCAAs, circadian rhythms and sleep disturbances

Given their involvement in peripheral and brain metabolism, are BCAAs related to circadian rhythms and sleep? It has long been observed that plasma concentrations of BCAAs and other essential amino acids oscillate over a 24h period [57,58]. Metabolomic studies have largely confirmed the robustness of these rhythms [59-64] but it should be noted that one of these analyses did not detect a rhythmicity of BCAAs [65], and a recent study comparing group to individual data drew attention to the fact that oscillations could be detected at the group level and not necessarily in each of its members, and conversely clear rhythms in some individuals are not detectable when the group is considered [66]. Plasma BCAA concentrations increase during the night at very similar times in humans and mice even though the latter are nocturnal [60,61]. Plasma BCAAs have thus been included in the list of metabolites that allow the calculation of circadian clock parameters [60,61]. Their cyclicity is indeed maintained under constant routine conditions, indicating that it is independent of food intake, physical activity, and sleep-wake cycles. A study in mice examining circadian fluctuations of amino acids in different tissues confirmed a possible link between the cyclicity of plasma BCAAs and those of free BCAAs present in muscles [59]. Interestingly, this study also showed that a high-fat diet, a risk factor for obesity, abolishes the circadian rhythms of muscle BCAAs. This phenomenon could contribute to the increase in plasma BCAAs induced by a fatty diet and robustly associated with obesity and diabetes [67,68]. It has been proposed that elevated plasma BCAAs are responsible for insulin resistance and hyperglycemia, but the mechanisms involved are still hypothetical [69]. Consistent with these observations, KO inactivation of the Bmal gene in the skeletal muscle has a major impact on lipid metabolism and on that of amino acids, including BCAAs [70]. A study in these mice even indicates that the muscle peripheral clock is necessary and sufficient to regulate certain aspects of the sleep-wake cycle, including its homeostatic regulation [71]. Outside of the muscles, there does not appear to be a clear correlation between plasma BCAAs and tissue BCAAs, particularly those in the prefrontal cortex and the master circadian clock in the suprachiasmatic nucleus [59].

Metabolomic explorations in humans and animals have also identified BCAAs as being impacted by sleep disturbances [72,73]. A study conducted in insomniac patients revealed changes depending on the time of day and the amino acid considered, as well as shifts in rhythms [62]: thus leucine and isoleucine are reduced during the night only, whereas valine is higher whatever the phase considered. This study also detected variations restricted to the night period in the concentrations of BCAA degradation metabolites. In a large population of healthy individuals, a study based on self-reported sleep data showed an association between bedtime and higher concentrations of the three BCAAs [74]. Sleep deprivation or sleep restriction in healthy subjects affects one or more BCAAs [68,75], in some but not all studies [72]. In sleep-deprived or sleep-restricted rats, an increase in leucine and valine was also noted in two metabolomic analyses [76,77]. This increase appears to be specific to plasma and is not observed at the level of the different brain areas, as was already the case for the circadian rhythmicity of BCAAs [76,78]. It should be noted that the three BCAAs are not all affected in the same way in these studies, which probably reflects specificities in their regulation.

Variations in BCAA concentration could impact sleep or even the circadian clock, outside of any disturbance or pathology. This was recently suggested for tryptophan in mice fed a diet deficient in particular amino acids over two weeks, with effects on the amplitude of locomotor rhythms [79]. A diet without BCAAs was without effect on the clock in this study, but it should be kept in mind that a deficit of amino acids in the diet can be at least partially compensated for, e.g., by a higher release from muscle tissue. On the other hand, a beneficial effect of BCAA supplementation on sleep has been demonstrated in a mouse model of brain injury [80]. In this model, cognitive deficits and decreased excitability in hippocampal networks related to a local reduction of BCAA (see previous section [48]), are also associated with sleep fragmentation, drowsiness, and decreased activity of orexinergic neurons. BCAA supplementation at least partially corrects all these parameters, suggesting that orexinergic neurons are responsible for the arousal deficits in this context. Orexinergic neurons are activated by exposure to non-essential amino acids but not by BCAAs [81]. However, BCAAs can impact the concentrations of non-essential amino acids such as glutamate and glutamine (see above). Despite all these studies, the origin of plasma BCAA cyclicity remains mysterious as well as their differential regulation following disturbed sleep or clock.

Plasma BCAA rhythmicity is remarkably conserved in humans, mice and rats and probably in other mammals, but is it a more generally conserved feature within the animal kingdom, and in particular in Drosophila, the model that facilitated the identification of the majority of the molecular components of the circadian clock [82]?

Drosophila, a model for the study of BCAAs?

Amino acid metabolism has not been as well studied in Drosophila as it has been in mammals. However, the many molecular genetic tools available in Drosophila, a model intensively studied not only for the circadian clock, but also for the sleep-wake cycle, memory, neurodegenerative diseases, metabolism and aging to name a few, make it indeed an extremely versatile experimental system to answer many questions [83]. As in mammals, it has been shown in Drosophila that leucine can trigger insulin secretion via glutamate dehydrogenase, involving JhI-21 and Minidisc, two LAT-1-like transporters [84]. The amino acid requirement kinases GCN2 and TOR play a similar role in Drosophila [85]. A recent study showed that Drosophila could detect leucine deficiency in the diet through sestrin expressed in glial cells, adding another essential amino acid sensor conserved between insects and mammals [86]. Enzymes for BCAA metabolism are also conserved, indicating that these metabolic pathways are comparable. A nuclear magnetic resonance assessment of free amino acids in whole Drosophila showed a circadian rhythmicity for BCAAs with a midnight peak remarkably similar to that observed in mammalian plasma [87]. However, a study by another laboratory using mass spectrometry did not replicate this result [88]. Further, more refined studies are needed to confirm these data. Quantitative assessment of circulating amino acids in hemolymph, the equivalent of blood in Drosophila, is not easy: because of the small size of the animal, it requires sacrificing it to obtain a measurement. Nevertheless, this analysis is possible, and could allow the identification of circadian oscillations [89,90]. Although this area of research is still emerging in Drosophila, intriguing data suggest that the metabolism of essential amino acids and in particular BCAAs may play a crucial role in the regulation of wakefulness/sleep. An example is given by the study of the mutant for GABA transaminase (GABAT)[91]. GABAT catalyzes the degradation of GABA by generating glutamate and succinic semialdehyde, both of which can feed the Krebs cycle and energy metabolism. Because of the lack of a degradation pathway, GABA levels are doubled in the brain, and the mutant flies sleep 2-3 hours longer than their genetic controls. When the mutants are placed on a food without amino acids (a sugar agar), GABA levels increase even more (600%, compared to 200% with a complete food) and the flies die within a few days, whereas wild-type flies can survive under the same conditions without apparent problems for 4 to 5 weeks [92]. Normal viability can be restored in mutants by adding only glutamate, or BCAAs (leucine and valine) whose transamination can provide glutamate in the cells, to the sugar agar. These data indicate that GABA degradation and BCAAs as a source of glutamate play a substantial role in brain energy metabolism. Another study conducted in our laboratory highlighted the role of the LAT-1-like transporters, JhI-21 and Minidisc, in the regulation of wakefulness and sleep [93]. Inhibition of the expression of these two genes in dopaminergic neurons increases sleep, especially during the night, and the opposite effect is observed when their expression is inhibited in GABAergic neurons. On the other hand, our recent work shows that inhibition of the expression of these two genes in the surface glia, the equivalent of the blood-brain barrier, shifts and fragments sleep at the beginning of the night, in connection with a reduction in the levels of GABA release [94]. This sleep phenotype can be normalized with leucine supplementation. These data suggest that BCAA transport at the beginning of the primary sleep phase is required to enhance the GABAergic transmission necessary for sleep initiation. This transport is also modulated by genes of the BLOC-1 complex, an endosomal complex whose loss of function is linked to schizophrenia [95], a psychiatric disorder associated with sleep disorders [96]. Thus, new avenues of research are emerging from the use of the Drosophila model, which could be tested in mammals.

Perspectives

In conclusion, BCAAs are metabolites of great interest for understanding the interactions between sleep, circadian clock and metabolism. Ongoing research may lead to the development of therapeutic strategies: for example, the study of BCAAs in traumatic brain injury in mice has shown a beneficial effect of BCAA supplementation on sleep/wake and this is being successfully tested in patients with similar lesions [97]. Data on LAT-1-like transporters in Drosophila surface glia indicate that BCAAs play an important role in sleep onset and therapeutic approaches could be developed based on this concept [94]. Because of the robust rhythmicity of plasma BCAAs and their competition with each other and with other amino acids for uptake by cells, it is important to optimize timing, composition and concentration in this type of approach. The mystery of the mechanisms of BCAA rhythmicity is still far from being solved. Multi-organ metabolomic analyses have advanced our knowledge, but studies at the cellular level using genetically encoded sensors [98] and the use of models such as Drosophila would allow for an even more detailed analysis of metabolic fluxes and their role.

**Acknowledgements**

I thank the French Society of Sleep Research and Medicine for its financial support to the research projects of my laboratory mentioned in this text [93,94]. The CNRS, INSERM, Claude Bernard

University Lyon1, also financially support this research. Thanks to Claude Gronfier for his critical reading of the manuscript.

**References**

[1] Oesch LT, Adamantidis AR. Sleep and Metabolism: Implication of Lateral Hypothalamic Neurons. Front Neurol Neurosci 2021;45:75–90. https://doi.org/10.1159/000514966.

[2] Spiegel K, Tasali E, Leproult R, Van Cauter E. Effects of poor and short sleep on glucose metabolism and obesity risk. Nat Rev Endocrinol 2009;5:253–61. https://doi.org/10.1038/nrendo.2009.23.

[3] Inocente CO, Lavault S, Lecendreux M, Dauvilliers Y, Reimao R, Gustin M-P, et al. Impact of obesity in children with narcolepsy. CNS Neurosci Ther 2013;19:521–8. https://doi.org/10.1111/cns.12105.

[4] Brosnan JT, Brosnan ME. Branched-Chain Amino Acids: Enzyme and Substrate Regulation. The Journal of Nutrition 2006;136:207S-211S. https://doi.org/10.1093/jn/136.1.207S.

[5] Bogatyreva NS, Finkelstein AV, Galzitskaya OV. Trend of amino acid composition of proteins of different taxa. J Bioinform Comput Biol 2006;4:597–608. https://doi.org/10.1142/s0219720006002016.

[6] Gardini S, Cheli S, Baroni S, Di Lascio G, Mangiavacchi G, Micheletti N, et al. On Nature's Strategy for Assigning Genetic Code Multiplicity. PLoS One 2016;11:e0148174. https://doi.org/10.1371/journal.pone.0148174.

[7] Abumrad NN, Miller B. The physiologic and nutritional significance of plasma-free amino acid levels. JPEN J Parenter Enteral Nutr 1983;7:163–70. https://doi.org/10.1177/0148607183007002163.

[8] Fernstrom JD, Wurtman RJ, Hammarstrom-Wiklund B, Rand WM, Munro HN, Davidson CS. Diurnal variations in plasma concentrations of tryptophan, tryosine, and other neutral amino acids: effect of dietary protein intake. Am J Clin Nutr 1979;32:1912–22. https://doi.org/10.1093/ajcn/32.9.1912.

[9] Frame EG. The Levels of Individual Free Amino Acids in the Plasma of Normal Man at Various Intervals After a High-Protein Meal1. J Clin Invest 1958;37:1710–23.

[10] Elovaris RA, Hutchison AT, Lange K, Horowitz M, Feinle-Bisset C, Luscombe-Marsh ND. Plasma Free Amino Acid Responses to Whey Protein and Their Relationships with Gastric Emptying, Blood Glucose- and Appetite-Regulatory Hormones and Energy Intake in Lean Healthy Men. Nutrients 2019;11:2465. https://doi.org/10.3390/nu11102465.

[11] Luscombe-Marsh ND, Hutchison AT, Soenen S, Steinert RE, Clifton PM, Horowitz M, et al. Plasma Free Amino Acid Responses to Intraduodenal Whey Protein, and Relationships with Insulin, Glucagon-Like Peptide-1 and Energy Intake in Lean Healthy Men. Nutrients 2016;8:4. https://doi.org/10.3390/nu8010004.

[12] Yang J, Chi Y, Burkhardt BR, Guan Y, Wolf BA. Leucine metabolism in regulation of insulin secretion from pancreatic beta cells. Nutr Rev 2010;68:270–9. https://doi.org/10.1111/j.1753-4887.2010.00282.x.

[13] Grill V, Björkman O, Gutniak M, Lindqvist M. Brain uptake and release of amino acids in nondiabetic and insulin-dependent diabetic subjects: important role of glutamine release for nitrogen balance. Metabolism 1992;41:28–32. https://doi.org/10.1016/0026-0495(92)90186-e.

[14] Yoshizawa F. Regulation of protein synthesis by branched-chain amino acids in vivo. Biochemical and Biophysical Research Communications 2004;313:417–22. https://doi.org/10.1016/j.bbrc.2003.07.013.

[15] Adibi SA. Metabolism of branched-chain amino acids in altered nutrition. Metabolism 1976;25:1287–302. https://doi.org/10.1016/S0026-0495(76)80012-1.

[16] Felig P, Owen OE, Wahren J, Cahill GF. Amino acid metabolism during prolonged starvation. J Clin Invest 1969;48:584–94. https://doi.org/10.1172/JCI106017.

[17] Bröer S, Bröer A. Amino acid homeostasis and signalling in mammalian cells and organisms. Biochem J 2017;474:1935–63. https://doi.org/10.1042/BCJ20160822.
[18] Metcoff J. Intracellular amino acid levels as predictors of protein synthesis. J Am Coll Nutr 1986;5:107–20. https://doi.org/10.1080/07315724.1986.10720118.
[19] Johnson C, Metcoff J. Relation of protein synthesis to plasma and cell amino acids in neonates. Pediatr Res 1986;20:140–6. https://doi.org/10.1203/00006450-198602000-00009.
[20] Herbert JD, Coulson RA, Hernandez T. Free amino acids in the caiman and rat. Comp Biochem Physiol 1966;17:583–98. https://doi.org/10.1016/0010-406x(66)90589-5.
[21] Bergström J, Fürst P, Norée LO, Vinnars E. Intracellular free amino acid concentration in human muscle tissue. J Appl Physiol 1974;36:693–7. https://doi.org/10.1152/jappl.1974.36.6.693.
[22] Boado RJ, Li JY, Nagaya M, Zhang C, Pardridge WM. Selective expression of the large neutral amino acid transporter at the blood–brain barrier. Proc Natl Acad Sci U S A 1999;96:12079–84.
[23] Scalise M, Galluccio M, Console L, Pochini L, Indiveri C. The Human SLC7A5 (LAT1): The Intriguing Histidine/Large Neutral Amino Acid Transporter and Its Relevance to Human Health. Front Chem 2018;6:243. https://doi.org/10.3389/fchem.2018.00243.
[24] Errasti-Murugarren E, Palacín M. Heteromeric Amino Acid Transporters in Brain: from Physiology to Pathology. Neurochem Res 2022;47:23–36. https://doi.org/10.1007/s11064-021-03261-w.
[25] Smith QR, Momma S, Aoyagi M, Rapoport SI. Kinetics of Neutral Amino Acid Transport Across the Blood-Brain Barrier. Journal of Neurochemistry 1987;49:1651–8. https://doi.org/10.1111/j.1471-4159.1987.tb01039.x.
[26] Tarlungeanu DC, Deliu E, Dotter CP, Kara M, Janiesch PC, Scalise M, et al. Impaired amino acid transport at the blood brain barrier is a cause of autism spectrum disorder. Cell 2016;167:1481-1494.e18. https://doi.org/10.1016/j.cell.2016.11.013.
[27] Napolitano L, Scalise M, Galluccio M, Pochini L, Albanese LM, Indiveri C. LAT1 is the transport competent unit of the LAT1/CD98 heterodimeric amino acid transporter. Int J Biochem Cell Biol 2015;67:25–33. https://doi.org/10.1016/j.biocel.2015.08.004.
[28] Dolgodilina E, Camargo SM, Roth E, Herzog B, Nunes V, Palacín M, et al. Choroid plexus LAT2 and SNAT3 as partners in CSF amino acid homeostasis maintenance. Fluids Barriers CNS 2020;17:1–12. https://doi.org/10.1186/s12987-020-0178-x.
[29] Pardridge WM. Blood-brain barrier carrier-mediated transport and brain metabolism of amino acids. Neurochem Res 1998;23:635–44. https://doi.org/10.1023/a:1022482604276.
[30] Oldendorf W. Brain uptake of radiolabeled amino acids, amines, and hexoses after arterial injection. American Journal of Physiology-Legacy Content 1971;221:1629–39. https://doi.org/10.1152/ajplegacy.1971.221.6.1629.
[31] Novarino G, El-Fishawy P, Kayserili H, Meguid NA, Scott EM, Schroth J, et al. Mutations in BCKD-kinase Lead to a Potentially Treatable Form of Autism with Epilepsy. Science 2012;338:394–7. https://doi.org/10.1126/science.1224631.
[32] Fernstrom JD. Diet-induced changes in plasma amino acid pattern: effects on the brain uptake of large neutral amino acids, and on brain serotonin synthesis. J Neural Transm Suppl 1979:55–67. https://doi.org/10.1007/978-3-7091-2243-3_5.
[33] Choi S, Disilvio B, Fernstrom MH, Fernstrom JD. Oral branched-chain amino acid supplements that reduce brain serotonin during exercise in rats also lower brain catecholamines. Amino Acids 2013;45:1133–42. https://doi.org/10.1007/s00726-013-1566-1.
[34] Fernstrom JD. Large neutral amino acids: dietary effects on brain neurochemistry and function. Amino Acids 2013;45:419–30. https://doi.org/10.1007/s00726-012-1330-y.
[35] Fernstrom JD, Fernstrom MH. Tyrosine, phenylalanine, and catecholamine synthesis and function in the brain. J Nutr 2007;137:1539S-1547S; discussion 1548S. https://doi.org/10.1093/jn/137.6.1539S.
[36] Hartmann EL. Effect of l-Tryptophan and Other Amino Acids on Sleep. Nutrition Reviews 1986;44:70–3. https://doi.org/10.1111/j.1753-4887.1986.tb07680.x.
[37] Saidi O, Rochette E, Doré É, Maso F, Raoux J, Andrieux F, et al. Randomized Double-Blind Controlled Trial on the Effect of Proteins with Different Tryptophan/Large Neutral Amino Acid

Ratios on Sleep in Adolescents: The PROTMORPHEUS Study. Nutrients 2020;12:E1885. https://doi.org/10.3390/nu12061885.

[38] Yudkoff M. Interactions in the Metabolism of Glutamate and the Branched-Chain Amino Acids and Ketoacids in the CNS. Neurochem Res 2017;42:10–8. https://doi.org/10.1007/s11064-016-2057-z.

[39] Sperringer JE, Addington A, Hutson SM. Branched-Chain Amino Acids and Brain Metabolism. Neurochem Res 2017;42:1697–709. https://doi.org/10.1007/s11064-017-2261-5.

[40] Hull J, Hindy ME, Kehoe PG, Chalmers K, Love S, Conway ME. Distribution of the branched chain aminotransferase proteins in the human brain and their role in glutamate regulation. J Neurochem 2012;123:997–1009. https://doi.org/10.1111/jnc.12044.

[41] Sakai R, Cohen DM, Henry JF, Burrin DG, Reeds PJ. Leucine-nitrogen metabolism in the brain of conscious rats: its role as a nitrogen carrier in glutamate synthesis in glial and neuronal metabolic compartments. J Neurochem 2004;88:612–22. https://doi.org/10.1111/j.1471-4159.2004.02179.x.

[42] Bak LK, Johansen ML, Schousboe A, Waagepetersen HS. Valine but not leucine or isoleucine supports neurotransmitter glutamate synthesis during synaptic activity in cultured cerebellar neurons. J Neurosci Res 2012;90:1768–75. https://doi.org/10.1002/jnr.23072.

[43] Sweatt AJ, Garcia-Espinosa MA, Wallin R, Hutson SM. Branched-chain amino acids and neurotransmitter metabolism: Expression of cytosolic branched-chain aminotransferase (BCATc) in the cerebellum and hippocampus. Journal of Comparative Neurology 2004;477:360–70. https://doi.org/10.1002/cne.20200.

[44] Zeisel A, Muñoz-Manchado AB, Codeluppi S, Lönnerberg P, La Manno G, Juréus A, et al. Cell types in the mouse cortex and hippocampus revealed by single-cell RNA-seq. Science 2015;347:1138–42. https://doi.org/10.1126/science.aaa1934.

[45] Dolgodilina E, Imobersteg S, Laczko E, Welt T, Verrey F, Makrides V. Brain interstitial fluid glutamine homeostasis is controlled by blood–brain barrier SLC7A5/LAT1 amino acid transporter. J Cereb Blood Flow Metab 2016;36:1929–41. https://doi.org/10.1177/0271678X15609331.

[46] Zaragozá R. Transport of Amino Acids Across the Blood-Brain Barrier. Front Physiol 2020;11. https://doi.org/10.3389/fphys.2020.00973.

[47] Wang L, Tu P, Bonet L, Aubrey KR, Supplisson S. Cytosolic Transmitter Concentration Regulates Vesicle Cycling at Hippocampal GABAergic Terminals. Neuron 2013;80:143–58. https://doi.org/10.1016/j.neuron.2013.07.021.

[48] Cole JT, Mitala CM, Kundu S, Verma A, Elkind JA, Nissim I, et al. Dietary branched chain amino acids ameliorate injury-induced cognitive impairment. Proc Natl Acad Sci U S A 2010;107:366–71. https://doi.org/10.1073/pnas.0910280107.

[49] Takeuchi F, Nishikata N, Nishimura M, Nagao K, Kawamura M. Leucine-Enriched Essential Amino Acids Enhance the Antiseizure Effects of the Ketogenic Diet in Rats. Front Neurosci 2021;15:637288. https://doi.org/10.3389/fnins.2021.637288.

[50] Evangeliou A, Spilioti M, Doulioglou V, Kalaidopoulou P, Ilias A, Skarpalezou A, et al. Branched Chain Amino Acids as Adjunctive Therapy to Ketogenic Diet in Epilepsy: Pilot Study and Hypothesis. J Child Neurol 2009;24:1268–72. https://doi.org/10.1177/0883073809336295.

[51] García-Espinosa MA, Wallin R, Hutson SM, Sweatt AJ. Widespread neuronal expression of branched-chain aminotransferase in the CNS: implications for leucine/glutamate metabolism and for signaling by amino acids. Journal of Neurochemistry 2007;100:1458–68. https://doi.org/10.1111/j.1471-4159.2006.04332.x.

[52] Cole JT, Sweatt AJ, Hutson SM. Expression of mitochondrial branched-chain aminotransferase and α-keto-acid dehydrogenase in rat brain: implications for neurotransmitter metabolism. Front Neuroanat 2012;6:18. https://doi.org/10.3389/fnana.2012.00018.

[53] Hull J, Usmari Moraes M, Brookes E, Love S, Conway ME. Distribution of the branched-chain α-ketoacid dehydrogenase complex E1α subunit and glutamate dehydrogenase in the human brain and their role in neuro-metabolism. Neurochemistry International 2018;112:49–58. https://doi.org/10.1016/j.neuint.2017.10.014.

[54] Murín R, Hamprecht B. Metabolic and regulatory roles of leucine in neural cells. Neurochem Res 2008;33:279–84. https://doi.org/10.1007/s11064-007-9444-4.
[55] Kato T, Pothula S, Liu R-J, Duman CH, Terwilliger R, Vlasuk GP, et al. Sestrin modulator NV-5138 produces rapid antidepressant effects via direct mTORC1 activation. J Clin Invest n.d.;129:2542–54. https://doi.org/10.1172/JCI126859.
[56] Leib DE, Knight ZA. Re-examination of Dietary Amino Acid Sensing Reveals a GCN2-Independent Mechanism. Cell Rep 2015;13:1081–9. https://doi.org/10.1016/j.celrep.2015.09.055.
[57] Eriksson T, Voog L, Wålinder J, Eriksson TE. Diurnal rhythm in absolute and relative concentrations of large neutral amino acids in human plasma. J Psychiatr Res 1989;23:241–9. https://doi.org/10.1016/0022-3956(89)90029-0.
[58] Feigin RD, Klainer AS, Beisel WR. Factors affecting circadian periodicity of blood amino acids in man. Metabolism 1968;17:764–75. https://doi.org/10.1016/0026-0495(68)90026-7.
[59] Dyar KA, Lutter D, Artati A, Ceglia NJ, Liu Y, Armenta D, et al. Atlas of Circadian Metabolism Reveals System-wide Coordination and Communication between Clocks. Cell 2018;174:1571-1585.e11. https://doi.org/10.1016/j.cell.2018.08.042.
[60] Minami Y, Kasukawa T, Kakazu Y, Iigo M, Sugimoto M, Ikeda S, et al. Measurement of internal body time by blood metabolomics. Proc Natl Acad Sci U S A 2009;106:9890–5. https://doi.org/10.1073/pnas.0900617106.
[61] Kasukawa T, Sugimoto M, Hida A, Minami Y, Mori M, Honma S, et al. Human blood metabolite timetable indicates internal body time. Proc Natl Acad Sci U S A 2012;109:15036–41. https://doi.org/10.1073/pnas.1207768109.
[62] Gehrman P, Sengupta A, Harders E, Ubeydullah E, Pack AI, Weljie A. Altered diurnal states in insomnia reflect peripheral hyperarousal and metabolic desynchrony: a preliminary study. Sleep 2018;41:zsy043. https://doi.org/10.1093/sleep/zsy043.
[63] Ang JE, Revell V, Anuska M, Mäntele S, Otway DT, Johnston JD, et al. Identification of Human Plasma Metabolites Exhibiting Time-of-Day Variation Using an Untargeted Liquid Chromatography–Mass Spectrometry Metabolomic Approach. Chronobiol Int 2012;29:868–81. https://doi.org/10.3109/07420528.2012.699122.
[64] Grant LK, Ftouni S, Nijagal B, De Souza DP, Tull D, McConville MJ, et al. Circadian and wake-dependent changes in human plasma polar metabolites during prolonged wakefulness: A preliminary analysis. Sci Rep 2019;9:4428. https://doi.org/10.1038/s41598-019-40353-8.
[65] Dallmann R, Viola AU, Tarokh L, Cajochen C, Brown SA. The human circadian metabolome. Proc Natl Acad Sci U S A 2012;109:2625–9. https://doi.org/10.1073/pnas.1114410109.
[66] Depner CM, Cogswell DT, Bisesi PJ, Markwald RR, Cruickshank-Quinn C, Quinn K, et al. Developing preliminary blood metabolomics-based biomarkers of insufficient sleep in humans. Sleep 2020;43:zsz321. https://doi.org/10.1093/sleep/zsz321.
[67] Bloomgarden Z. Diabetes and branched-chain amino acids: What is the link? J Diabetes 2018;10:350–2. https://doi.org/10.1111/1753-0407.12645.
[68] Bell LN, Kilkus JM, Booth JN, Bromley LE, Imperial JG, Penev PD. Effects of Sleep Restriction on the Human Plasma Metabolome. Physiol Behav 2013;0:10.1016/j.physbeh.2013.08.007. https://doi.org/10.1016/j.physbeh.2013.08.007.
[69] Adams SH. Emerging Perspectives on Essential Amino Acid Metabolism in Obesity and the Insulin-Resistant State12. Adv Nutr 2011;2:445–56. https://doi.org/10.3945/an.111.000737.
[70] Dyar KA, Hubert MJ, Mir AA, Ciciliot S, Lutter D, Greulich F, et al. Transcriptional programming of lipid and amino acid metabolism by the skeletal muscle circadian clock. PLoS Biol 2018;16:e2005886. https://doi.org/10.1371/journal.pbio.2005886.
[71] Ehlen JC, Brager AJ, Baggs J, Pinckney L, Gray CL, DeBruyne JP, et al. Bmal1 function in skeletal muscle regulates sleep. Elife 2017;6:e26557. https://doi.org/10.7554/eLife.26557.
[72] Humer E, Pieh C, Brandmayr G. Metabolomics in Sleep, Insomnia and Sleep Apnea. Int J Mol Sci 2020;21:E7244. https://doi.org/10.3390/ijms21197244.
[73] Malik DM, Paschos GK, Sehgal A, Weljie AM. Circadian and Sleep Metabolomics Across Species. Journal of Molecular Biology 2020;432:3578–610. https://doi.org/10.1016/j.jmb.2020.04.027.
[74] Xiao Q, Derkach A, Moore SC, Zheng W, Shu X-O, Gu F, et al. Habitual Sleep and human plasma metabolomics. Metabolomics 2017;13:63. https://doi.org/10.1007/s11306-017-1205-z.

[75] Skene DJ, Skornyakov E, Chowdhury NR, Gajula RP, Middleton B, Satterfield BC, et al. Separation of circadian- and behavior-driven metabolite rhythms in humans provides a window on peripheral oscillators and metabolism. Proceedings of the National Academy of Sciences 2018;115:7825–30. https://doi.org/10.1073/pnas.1801183115.

[76] Gou X, Cen F, Fan Z, Xu Y, Shen H, Zhou M. Serum and Brain Metabolomic Variations Reveal Perturbation of Sleep Deprivation on Rats and Ameliorate Effect of Total Ginsenoside Treatment. Int J Genomics 2017;2017:5179271. https://doi.org/10.1155/2017/5179271.

[77] Weljie AM, Meerlo P, Goel N, Sengupta A, Kayser MS, Abel T, et al. Oxalic acid and diacylglycerol 36:3 are cross-species markers of sleep debt. Proc Natl Acad Sci U S A 2015;112:2569–74. https://doi.org/10.1073/pnas.1417432112.

[78] Bourdon AK, Spano GM, Marshall W, Bellesi M, Tononi G, Serra PA, et al. Metabolomic analysis of mouse prefrontal cortex reveals upregulated analytes during wakefulness compared to sleep. Sci Rep 2018;8. https://doi.org/10.1038/s41598-018-29511-6.

[79] Petrus P, Cervantes M, Samad M, Sato T, Chao A, Sato S, et al. Tryptophan metabolism is a physiological integrator regulating circadian rhythms. Mol Metab 2022;64:101556. https://doi.org/10.1016/j.molmet.2022.101556.

[80] Lim MM, Elkind J, Xiong G, Galante R, Zhu J, Zhang L, et al. Dietary Therapy Mitigates Persistent Wake Deficits Caused by Mild Traumatic Brain Injury. Sci Transl Med 2013;5:215ra173. https://doi.org/10.1126/scitranslmed.3007092.

[81] Karnani MM, Apergis-Schoute J, Adamantidis A, Jensen LT, de Lecea L, Fugger L, et al. Activation of central orexin/hypocretin neurons by dietary amino acids. Neuron 2011;72:616–29. https://doi.org/10.1016/j.neuron.2011.08.027.

[82] Huang R-C. The discoveries of molecular mechanisms for the circadian rhythm: The 2017 Nobel Prize in Physiology or Medicine. Biomed J 2018;41:5–8. https://doi.org/10.1016/j.bj.2018.02.003.

[83] Bellen HJ, Tong C, Tsuda H. 100 years of Drosophila research and its impact on vertebrate neuroscience: a history lesson for the future. Nat Rev Neurosci 2010;11:514–22. https://doi.org/10.1038/nrn2839.

[84] Ziegler AB, Manière G, Grosjean Y. JhI-21 plays a role in Drosophila insulin-like peptide release from larval IPCs via leucine transport. Sci Rep 2018;8:1908. https://doi.org/10.1038/s41598-018-20394-1.

[85] Bjordal M, Arquier N, Kniazeff J, Pin JP, Léopold P. Sensing of Amino Acids in a Dopaminergic Circuitry Promotes Rejection of an Incomplete Diet in Drosophila. Cell 2014;156:510–21. https://doi.org/10.1016/j.cell.2013.12.024.

[86] Gu X, Jouandin P, Lalgudi PV, Binari R, Valenstein ML, Reid MA, et al. Sestrin mediates detection of and adaptation to low-leucine diets in Drosophila. Nature 2022;608:209–16. https://doi.org/10.1038/s41586-022-04960-2.

[87] Gogna N, Singh VJ, Sheeba V, Dorai K. NMR-based investigation of the Drosophila melanogaster metabolome under the influence of daily cycles of light and temperature. Mol BioSyst 2015;11:3305–15. https://doi.org/10.1039/C5MB00386E.

[88] Rhoades SD, Nayak K, Zhang SL, Sehgal A, Weljie AM. Circadian- and Light-driven Metabolic Rhythms in Drosophila melanogaster. J Biol Rhythms 2018;33:126–36. https://doi.org/10.1177/0748730417753003.

[89] Piyankarage SC, Augustin H, Featherstone DE, Shippy SA. Hemolymph amino acid variations following behavioral and genetic changes in individual Drosophila larvae. Amino Acids 2010;38:779–88. https://doi.org/10.1007/s00726-009-0284-1.

[90] Piyankarage SC, Augustin H, Grosjean Y, Featherstone DE, Shippy SA. Hemolymph amino acid analysis of individual Drosophila larvae. Anal Chem 2008;80:1201–7. https://doi.org/10.1021/ac701785z.

[91] Chen W-F, Maguire S, Sowcik M, Luo W, Koh K, Sehgal A. A neuron-glia interaction involving GABA transaminase contributes to sleep loss in sleepless mutants. Mol Psychiatry 2015;20:240–51. https://doi.org/10.1038/mp.2014.11.

[92] Maguire SE, Rhoades S, Chen W-F, Sengupta A, Yue Z, Lim JC, et al. Independent Effects of γ-Aminobutyric Acid Transaminase (GABAT) on Metabolic and Sleep Homeostasis. J Biol Chem 2015;290:20407–16. https://doi.org/10.1074/jbc.M114.602276.

[93] Aboudhiaf S, Alves G, Parrot S, Amri M, Simonnet MM, Grosjean Y, et al. LAT1-like transporters regulate dopaminergic transmission and sleep in Drosophila. Sleep 2018;41. https://doi.org/10.1093/sleep/zsy137.

[94] Li H, Aboudhiaf S, Parrot S, Scote-Blachon C, Benetollo C, Lin J-S, et al. Pallidin function in drosophila surface glia regulates sleep and is dependent on amino acid availability 2022:2022.05.03.490434. https://doi.org/10.1101/2022.05.03.490434.

[95] Ghiani CA, Dell'Angelica EC. Dysbindin-containing complexes and their proposed functions in brain: from zero to (too) many in a decade. ASN Neuro 2011;3. https://doi.org/10.1042/AN20110010.

[96] Ferrarelli F. Sleep disturbances in Schizophrenia and Psychosis. Schizophr Res 2020;221:1–3. https://doi.org/10.1016/j.schres.2020.05.022.

[97] Elliott JE, Keil AT, Mithani S, Gill JM, O'Neil ME, Cohen AS, et al. Dietary Supplementation With Branched Chain Amino Acids to Improve Sleep in Veterans With Traumatic Brain Injury: A Randomized Double-Blind Placebo-Controlled Pilot and Feasibility Trial. Front Syst Neurosci 2022;16:854874. https://doi.org/10.3389/fnsys.2022.854874.

[98] Yoshida T, Nakajima H, Takahashi S, Kakizuka A, Imamura H. OLIVe: A Genetically Encoded Fluorescent Biosensor for Quantitative Imaging of Branched-Chain Amino Acid Levels inside Single Living Cells. ACS Sens 2019;4:3333–42. https://doi.org/10.1021/acssensors.9b02067.

Figure 1

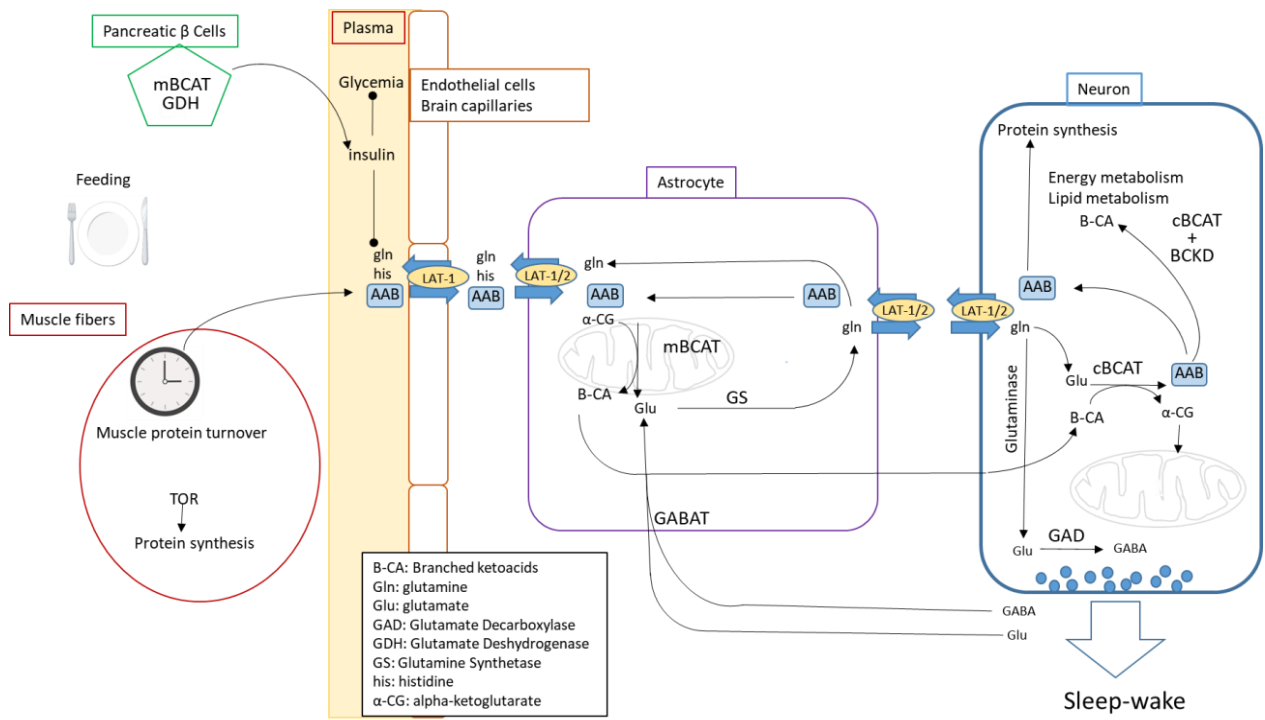

**Schematic model of BCAA metabolism.** A very simplified diagram of the elements described in the text. The BCAAs supplied by the diet are among the most rapidly transferred amino acids to the bloodstream. Outside of the diet, a portion of plasma BCAA's are derived from the turnover of muscle proteins, the body's primary amino acid pool, probably under the influence of the circadian clock. The peripheral clock of muscle fibers has a major impact on metabolism and sleep. BCAAs, especially leucine, stimulate protein synthesis, either directly or by activating TOR signaling. They also stimulate insulin secretion by the β-cells of the pancreas. In turn, insulin negatively regulates blood glucose and plasma amino acids. The LAT-1 transporter in blood-brain barrier endothelial cells allows entry of BCAAs into the brain parenchyma in exchange for glutamine and/or histidine. BCAAs are transferred to astrocytes via the possible transporters LAT-1 and LAT-2: they are transaminated by mitochondrial BCAT (mBCAT) and allow the generation of glutamate and a branched keto acid (B-CA). Glutamine synthetase (GS) converts glutamate to glutamine which can be exported in exchange for BCAA and/or transferred to neurons, again via the LAT-1 and/or LAT-2 transporters. In the neuron, glutaminase regenerates glutamate from glutamine and glutamate decarboxylase (GAD) converts glutamate to GABA in inhibitory interneurons. Branched ketoacid (B-CA) produced in astrocytes is also transferred to neurons where cytoplasmic BCAT (cBCAT) reconstitutes BCAA by consuming glutamate. These BCAAs can be used for protein synthesis, energy metabolism, lipid metabolism or transferred to astrocytes. By contributing to glutamate and GABA synthesis, BCAAs may play a significant role in the excitation/inhibition balance and in the regulation of wakefulness and sleep.